\documentclass[10pt,journal,compsoc]{IEEEtran}
\pdfoutput=1

\ifCLASSOPTIONcompsoc
  \usepackage[nocompress]{cite}
\else
  \usepackage{cite}
\fi
\ifCLASSINFOpdf
  \usepackage[pdftex]{graphicx}
\else
\fi
\usepackage{amsmath}

\usepackage{multirow}

\begin{document}
%
\title{Explainable and Lightweight Model for COVID-19 Detection Using Chest Radiology Images}
%
%
%
%

\author{Suba~S
        and~Nita~Parekh
\IEEEcompsocitemizethanks{\IEEEcompsocthanksitem Suba S and N. Parekh are with CCNSB, International Institute of Information Technology, Prof. C R Rao Road, Hyderabad, India 500032.\protect\\
E-mail: suba.s@research.iiit.ac.in, nita@iiit.ac.in
}
\thanks{.}}

%
%

\markboth{}%
{Shell \MakeLowercase{\textit{et al.}}: Bare Demo of IEEEtran.cls for Computer Society Journals}
%



\IEEEtitleabstractindextext{%
\begin{abstract}
Deep learning (DL) analysis of Chest X-ray (CXR) and Computed tomography (CT) images has garnered a lot of attention in recent times due to the COVID-19 pandemic. Convolutional Neural Networks (CNNs) are well suited for the image analysis tasks when trained on humongous amounts of data. Applications developed for medical image analysis require high sensitivity and precision compared to any other fields. Most of the tools proposed for detection of COVID-19 claims to have high sensitivity and recalls but have failed to generalize and perform when tested on unseen datasets. This encouraged us to develop a CNN model, analyze and understand the performance of it by visualizing the predictions of the model using class activation maps generated using (Gradient-weighted Class Activation Mapping) Grad-CAM technique. This study provides a detailed discussion of the success and failure of the proposed model at an image level. Performance of the model is compared with state-of-the-art DL models and shown to be comparable. The data and code used are available at https://github.com/aleesuss/c19.
\end{abstract}

}

\maketitle

\IEEEdisplaynontitleabstractindextext

%
\IEEEpeerreviewmaketitle

\IEEEraisesectionheading{\section{Introduction}\label{sec:introduction}}

%
%
%
%
\IEEEPARstart{E}{arly} diagnosis of coronavirus and effective prevention of transmission by isolating the infected individuals have been the core tasks in managing the COVID-19 pandemic. Shortage of RT-PCR kits, longer times required for the results, and high false negative rates during the early phases of infection have been a bottleneck faced worldwide for timely action. Chest radiography imaging (chest X-rays (CXRs) and computed tomography (CT) scans) were used in hospitals worldwide for faster triaging of patients and several studies have reported the advantages of using chest radiographs along with RT-PCR to confirm the diagnosis during the early phase of pandemic~\cite{xie2020chest},~\cite{cleverley2020role}. Earlier work on other strains of the coronavirus family, the Middle East respiratory syndrome (MERS) and severe acute respiratory syndrome (SARS) outbreaks also confirm the usefulness of chest radiographs in the diagnosis of pulmonary diseases~\cite{hosseiny2020radiology}. It has been observed that COVID patients exhibit changes in their chest radiographs even before the onset of other symptoms~\cite{shi2020radiological} making them a useful diagnostic tool for early diagnosis. Since machine learning and artificial intelligence are well established methods in image analysis, there has been much attention in developing deep learning (DL) models for CXR and CT image analysis for the detection of COVID-19. Some characteristic features associated with COVID-19 include ground-glass opacification (GGO), consolidation, and crazy lines~\cite{lee2020covid},~\cite{litmanovich2020review}. Of these, the most common feature, GGOs, are not easily perceivable posing challenges. Further, large inflow of patients during various waves of pandemic worldwide has caused heavy burden on the hospital staff.  Thus, automated and accurate detection of the disease is desirable during such situations. Artificial Intelligence (AI) assisted diagnosis would prove efficient as machine learning models are good at capturing subtle patterns in data and are quick in arriving at results with expert level performances. But, despite the excellent performance of machine learning models their applicability in healthcare has been limited due to its ‘black box’ nature. For these tools to gain acceptance in healthcare systems, there is need for the models to be interpretable and explainable. In this paper we propose a simple diagnostic tool built using convolution neural network (CNN) which can take either chest X-ray (CXR) or chest computed tomography (CT) images of a person as input for the detection of COVID-19. Here we present the performance of the model for both binary classification (i.e., COVID-19 vs non-Covid) or three-class classification (Normal, COVID-19, or non-Covid pneumonia). To make the predictions of the model interpretable, visualization of the image with the heatmap is provided. This feature highlights the pixels in the image that correspond to COVID-19 features based on which the decision was made by the model. The dataset used for training the models are the largest of its kind that are openly available. The models’ performance is evaluated with other state-of-the-art deep learning models such as VGG-16, ResNet-50 and Inception-v3. The performance of the lightweight CNN is comparable to very deep  models with much fewer number of parameters and smaller training time, making it suitable for clinical settings. The data and code are available at https://github.com/aleesuss/c19.

\section{Related Works}
Like with any pneumonia, COVID-19 also causes the lung density to increase, seen as whiteness in the lungs in chest radiography images. Based on disease severity, the lung markings in the images exhibit different patterns; when partially obscured it is called a ground glass pattern and when completely obscured by the whiteness, it is called consolidation~\cite{cleverley2020role}. Chest CTs and CXRs have been widely used in diagnosis of pneumonia. Depending on the availability of the tools, any of these image modalities could be used for identifying lung abnormalities as both of them capture the lung features in COVID-19 pneumonia such as Ground Glass Opacities (GGO), consolidations with or without vascular enlargement, interlobular septal thickening, air bronchogram sign, etc. with multiple GGOs being the common of them all mostly in the middle and lower lobes of lungs~\cite{li2020coronavirus}. Identification of these subtle patterns result in variations among radiologists and burdens the experts in pandemic like situations. Application of artificial intelligence (AI) and explainable models can aid in addressing both these issues. Various studies have reported applications of deep learning models on radiography images in the diagnosis of diseases such as cancer, pulmonary diseases, cardiac diseases, etc.~\cite{litjens2017survey}. Many deep learning based models have also been developed to assist in the diagnosis of COVID-19 using chest radiology images. These studies have either used pre-defined architectures such as ResNet, VGG, Exception, UNet, etc. pre-trained on ImageNet database~\cite{deng2009imagenet} or have synthesized very complex models on these deep layered architectures. COVID-Net~\cite{wang2020covid} which is one of the first and most popular models to classify CXRs into Normal, Pneumonia and COVID-19 uses a tailor-made design pattern based on ResNet architecture that comprises projection-expansion-projection-extension (PEPX) pattern of multiple Conv layers generated using a machine driven Generator-Inquisitor pair to obtain an optimal model, pre-trained on ImageNet [9]. Oh et al.~\cite{oh2020deep} proposed DenseNet based model for segmenting the CXRs before classifying the images using ResNet-18 based classification models. The dataset used included a total of only 500 images for training, validation and testing of the model. A patch-based classification technique was proposed to overcome the small dataset problem where multiple patches were extracted from a single image and used for training multiple ResNet-18 models and majority voting taken to arrive at the final prediction, which was shown to have comparable performances to that of COVID-Net. The study by Ozturk et al.~\cite{ozturk2020automated} proposed a DarkNet-19 model, a CNN model with 19 convolutional layers to classify CXRs into binary and three-class classifications. Performance of DarkNet-19 model evaluated using heatmaps and assessed by expert radiologists showed agreement in identifying the affected regions in the images with an accuracy of 98\% for binary classification and 87\% for three-class classification. The model was trained and tested only on 1125 images including 125 COVID-19 images. Generative Adversarial Network (GAN) was used to generate CXR images from \begin{math}\sim300\end{math} original CXR images to overcome the limited dataset problem and improve model performances~\cite{loey2020within}. Alexnet, Googlenet and ResNet-18 were used for classification tasks in the study. 

COVIDNet-CT~\cite{gunraj2020covidnet} like COVID-Net uses the machine-driven design exploration strategy for building the model to classify CTs with ResNet type backbone architecture, and is also pretrained on ImageNet. The design exploration leverages generative synthesis to identify the network architecture by solving a constrained optimization problem strategy involving spatial, point-wise and depth-wise convolutions. ResNet-101 is found to have the best performance among 10 different CNN architectures in classifying CTs to COVID-19 and non-Covid classes using radiologist annotated image patches of infections in~\cite{ardakani2020application}. MobileNet V2 was used in classifying CT images to non-Covid and COVID-19 classes with shorter response time compared to VGG-16, VGG-19 and DenseNet 201~\cite{singh2022deep}. The performance validation of these models on hold-out test sets were more than 95\%. Wang et.al. uses InceptionNet architecture pretrained on ImageNet for classifying CTs into COVID-19 and other viral pneumonia classes~\cite{wang2021deep}. The model achieved 89\% accuracy on hold-out test set while an external test set gave 79\% accuracy with a sensitivity of 0.67. 

Maghdid et al.~\cite{maghdid2021diagnosing} proposed a CNN-based architecture for binary classification of COVID-19 positive and negative cases. Performance of the model was shown to be comparable to a pre-trained AlexNet model, with an accuracy of 94\% for both CXR and CT models using CNN and an accuracy of 98\% with CXR and 82\% with CT with AlexNet. In the study by Owais et al.~\cite{owais2021multilevel} CXR and CT images were combined for training the MobileNet model, which leverages both multi-level and multi-stage learning, and depth wise (DW) convolutions were used for reducing the computation costs in a standard convolution layer. Though an accuracy of 95\% was achieved by the model, the visualizations of the predictions were not convincing. This, according to the authors, was due to non-utilization of a well localized dataset, such as a segmented dataset or annotated dataset with abnormalities marked, for training. Numerous such studies have been published recently that considers both CXR and CT images for training the models and provide visualizations of predictions~\cite{panwar2020deep},~\cite{jia2021classification}.

Most of the earlier studies have recorded high sensitivity and specificity \begin{math}\sim99\%\end{math} in detecting COVID-19 with CT  images~\cite{ardakani2020application} and CXR images~\cite{zokaeinikoo2021aidcov} when evaluated on hold-out test sets, but as datasets have added more images, collected from across the world, and studies have utilized external datasets for performance evaluation of models, significant drop in performances are seen. High performances were recorded with binary classification of images (normal vs COVID-19) whereas three class (COVID-19, normal, pneumonia) and four classifications (COVID-19, normal, bacterial and viral pneumonia)~\cite{khan2020coronet},~\cite{mangal2020covidaid} were considerably low compared to binary classification performances . The major concern with these studies is that most of them fail to generalize and perform well when tested on external datasets, unseen by the models during training which limits their clinical utility~\cite{nguyen2021deep},~\cite{barish2021external}. This is due to the variability in distribution in the data, demographics of the patients used in training set, differences in image acquisition and reconstruction, etc. In the proposed study, an image level analysis of the visualization of results is carried out and the visualization maps are given that justify the performance of the proposed model by highlighting only the abnormal areas.

\section{Data}

\subsection{Chest X-Ray Dataset}
Chest X-Ray Dataset: The Chest X-Ray dataset considered in this study comprises collection of eight publicly available data repositories by~\cite{wang2020covid}, and is summarized in Table~\ref{table_1}. Of these, the NIHCC resource hosts only normal and pneumonia images, no COVID-19 image data. Data is prepared for both binary classification (COVID-19 vs non-Covid) and for three-class classification (COVID-19, normal and non-Covid pneumonia) tasks. The number of images for these two classification tasks are given in Table~\ref{table_2} and Table~\ref{table_3} respectively. The non-Covid class in the case of binary classification consists of both normal and other pneumonia images. The number of images is comparatively balanced in the two classes for the binary classification task, while for the three-class classification task, the COVID-19 class is \begin{math}\sim2\end{math}-3 times larger compared to the other two classes. However, it may be noted that while the number of images is same as number of patients in Normal and Pneumonia classes, COVID-19 class has a much larger number of 16,690 images from 2986 patients. Consequently, amount of variation the models see in the images for the two classes is much greater than in the COVID-19 class and is likely to compensate for data imbalance. All the images were resized to \begin{math}224 \times 224\end{math} pixels for training and testing. For the binary classification task, 200 COVID-19 and 200 non-Covid  images were randomly taken as test set and 80\% of the remaining images (16490 COVID-19, 13992 non-Covid) were used for training and 20\% for validation. For multi-class classification, the test set is same as the one used in binary classification, 100 images each in normal and pneumonia classes and 200 images in covid class. Training set consisted of 24,104 images and 6026 images for validation across the three classes. Test set consisted of 200 COVID images, and 100 images each for Normal and Pneumonia classes. 
\begin{table}[!t]
\renewcommand{\arraystretch}{1.3}
\caption{List of repositories used for collating chest X-ray and CT scan images for training and evaluating the proposed model}
\label{table_1}
\centering
\begin{tabular}{|c|c|}
\hline
 \bfseries Links for CXR Datasets & \bfseries Ref.\\
\hline
  github.com/ieee8023/covid-chestX-ray-dataset

&~\cite{cohen2020covid}\\
\hline
 github.com/agchung/Figure1-

& \\
 COVID-chestX-ray-dataset

&~\cite{wang2020covid}\\
\hline
 github.com/agchung/Actualmed-

&\\
 COVID-chestX-ray-dataset

&~\cite{wang2020covid}\\

\hline
kaggle.com/tawsifurrahman/covid19-

& \\
radiography-database

& --\\
\hline
 nihcc.app.box.com/v/ChestX-ray-NIHCC

&~\cite{wang2017chestx}\\
\hline
 wiki.cancerimagingarchive.net/pages/

& \\
viewpage.action?pageId=70230281

&~\cite{clark2013cancer}\\
\hline
wiki.cancerimagingarchive.net/pages/

& \\
 viewpage.action?pageId=89096912

&~\cite{clark2013cancer}\\
\hline
 bimcv.cipf.es/bimcv-projects/bimcv-covid19/

&~\cite{vaya2020bimcv}\\
\hline

\bfseries Links for CT Datasets & \bfseries Ref.\\
\hline
 CNCB (2019nCoVR) AI Diagnosis Dataset

&~\cite{zhang2020clinically}\\
\hline
 Lung CT Lesion Segmentation Challenge (COVID-19-20)

& --\\
\hline
 TCIA CT Images in COVID-19

&~\cite{an2020ct}\\
\hline
 COVID-19 CT Lung and Infection Segmentation Dataset

&~\cite{PPR:PPR346511}\\
\hline
 COVID-CTSet

&~\cite{rahimzadeh2020fully}\\
\hline
 Radiopaedia.org

&--\\
\hline
 LIDC-IDRI

&~\cite{armato2011lung}\\
\hline
 Integrative CT Images and Clinical Features for COVID-19

&--\\
\hline

\end{tabular}
\end{table}

\begin{table}
\renewcommand{\arraystretch}{1.3}
\caption{Training and test set data for binary classification using chest X-ray radiography images. Number of patients is given in brackets}
\label{table_2}
\centering
\begin{tabular}{|c|c|c|c|}
\hline
Type & non-Covid & COVID-19  & Total\\
\hline
Train & 13992 (13850) & 16490 (2808) & 30482(16648) \\
\hline
Test & 200 (200) & 200 (178) & 400 (378) \\
\hline
\end{tabular}
\end{table}

\begin{table}
\renewcommand{\arraystretch}{1.3}
\caption{Training and test set data for three-class classification using chest X-ray radiography images. Number of patients is given in brackets}
\label{table_3}
\centering
\begin{tabular}{|c|c|c|c|c|}
\hline
Type & Normal & Pneumonia & COVID-19 & Total\\
\hline
Train & 8085 (8085) & 5555 (5531) & 16490 (2808) & 30130 (16424) \\
\hline
Test & 100 (100) & 100 (100) & 200 (178) & 400 (378) \\
\hline
\end{tabular}
\end{table}

\begin{table}
\renewcommand{\arraystretch}{1.3}
\caption{The benchmark COVIDx-CT dataset considered for training, validation and testing for three-class classification using CT scan images is given. Number of patients is given in brackets.}
\label{table_4}
\centering
\begin{tabular}{|c|c|c|c|c|}
\hline
\bfseries Type &  \bfseries Normal & \bfseries Pneumonia & \bfseries Covid & \bfseries Total\\
\hline
    Train   & 35996  (321)  & 25496  (558)    & 82286  (1958)  & 143778  (2837)   \\
\hline
    Val     & 11842 (126)  & 7400 (190)     & 6244 (166)    & 25486 (482)      \\
\hline
    Test    & 12245 (126)  & 7395 (125)     & 6018 (175)    & 25658 (426)     \\
\hline
\end{tabular}
\end{table}

\subsection{Chest CT Dataset}
One of the largest publicly available benchmark dataset of chest CT images, COVIDx-CT, is used for training the CT-based model. It includes CT images of normal, pneumonia, and COVID-19 classes, collected from multiple data sources worldwide (given in Table~\ref{table_1}). It consists of 194922 CT images from 3745 patients of which 94,548 images are from 2299 COVID-19 patients with labels verified using RT-PCR or radiology reports. The dataset is split into 60-20-20 ratio for training, validation and testing the proposed model, as summarized in Table~\ref{table_4}. The split ratio was the same as in COVIDNet-CT study~\cite{gunraj2020covidnet}.

\begin{figure*}[!t]
\centering
\includegraphics[width=7.0in]{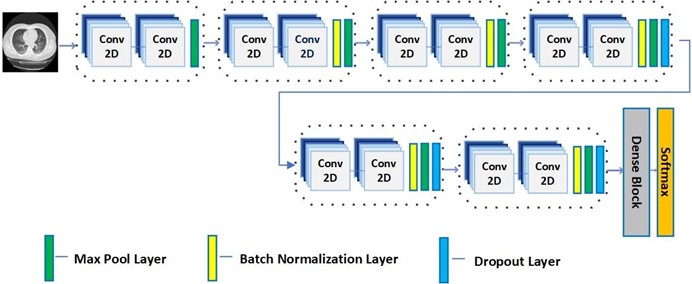}
\caption{Architecture of the proposed CNN model. The convolution blocks are marked with dotted rectangles with each block comprising two convolutional layers. The Max pool, batch normalization and dropout layers are colour coded as shown.}
\label{fig_1}
\end{figure*}

\section{Methods}
The X-ray machines are the most common imaging modality available even in primary healthcare centres and underdeveloped/rural areas. On the other hand, computed tomography (CT) machines are less accessible to majority of the population as these are expensive and available only in large healthcare facilities. Most importantly, radiation exposure is much less in the case of CXR compared to CT scan. Further, due to infection containment issues during patient transport to CT suites, decontamination of CT rooms after each patient, and lack of CT availability in many healthcare settings, portable chest radiography (CXR) has been proposed for early disease detection and follow up of lung abnormalities~\cite{jacobi2020portable}. Though less sensitive to CT scan imaging, portable CXR machines provide testing at patient bed, making them indispensable in triaging of COVID patients. Also, the results can be obtained in few minutes and has the capability of testing large inflow of patients in a short time. However, whenever available, CT scans are preferrable since Chest CT scans involve taking many X-ray measurements at different angles across a patient’s chest to produce cross-sectional images, thus providing a more sensitive testing. So, it is desirable that any automatic diagnostic tool should be able to handle both CXR and CT scan images. Considering these issues, in this study we propose a single light weight CNN model for both CXR and CT scan images.

CNNs have been widely used for medical image analysis not only in pulmonary diseases but various other cardio-thoracic conditions~\cite{mitra2020systematic} and simple CNN models are shown to have performances at par with very deep learning/ensemble models. We carried out a series of experiments to identify various hyperparameters for a most suitable model with the same architecture for the diagnosis of COVID-19 using either CXR or CT images. We started our experiments with CT images as CT dataset had a greater number of images. The basic architecture of the proposed CCN model is given in Fig.~\ref{fig_1}. The architecture only defers for the number of nodes in the output layer and the activation function used for binary and multiclass classification. It consists of 6 convolutional blocks with each block comprising two convolution layers. The first block has 16 filters followed by 32, 64, 128, 256, and 512 filters in successive blocks. All kernels are of size \begin{math} 3 \times 3 \end{math} and a zero padding is used to make the input and output width and height dimensions the same. ‘Maxpool’, ‘batchnormalization’ and ‘dropout’ layers are added to the model as shown in the figure. The convolutional layers are used for extracting the image features and ‘Maxpool’ layers down sample the image dimensions. ‘Dropout’ layers remove those neurons whose contribution to the output is low to reduce overfitting. The convolutional blocks are followed by dense layers with 512, 128, 64 and 1 (3 for multi class classification) nodes in each layer. Dropout layers are also used after each dense layer. Dense layers identify the non-linear combination of the extracted features to be used for final classification. The output layer has a ‘sigmoid’ (‘softmax' for multi class) activation function, and previous layers of convolution and dense layers used ‘Relu’ function. Loss function used was ‘binary cross entropy’ (‘categorical cross entropy’ for multi class). The input image dimensions are \begin{math} 224 \times 224 \times 3 \end{math}and batch size of 8 was used. 

The hyperparameters of the proposed model such as number of conv layers, dropout layers, number of epochs, etc. were chosen empirically. To find optimal number of convolution blocks, two experiments were performed using 5 \& 6-layered CNN architectures, repeated 3 times for each case and averaged results are given in Table 1 of supplementary file, S1. In the case of CT, though training accuracy increased from 97.8 to 99, validation accuracy dipped from 97.25 to 95.67, and test accuracy, 96.3, was comparable in the two experiments. This indicates the model may have started overfitting and we did not further increase the number of blocks. In the case of CXR also a similar trend was seen. Since adding a dense layer increased the number of parameters from 2.9M to 3.7M for the 6-layered architecture and the training accuracy was quite good \begin{math}(\sim 99)\end{math}, we did not play around with the number of dense layers. For choosing optimal number of epochs, 4 experiments were performed with 10, 18, 20, and 30 epochs for 6-layered CNN model and the results are given in Table 2 of supplementary file, S1. Number of nodes in first dense layer was chosen based on number of filters in previous Conv block and halved in subsequent dense layers. Number of filters in Conv blocks were increased hierarchically from 32 to 512. To select best validated model, after each epoch during training, performance of the model was evaluated on validation set and weights of the model were saved if accuracy of the model improved. Thus, after training, weights of the model that gave best accuracy across all the epochs were considered for testing phase. The results of these experiments are given in supplementary file, S1. 

CXRs and chest CTs are different image modalities of the same region and capture the same features of the infection, viz., GGOs, consolidations, etc.  It has been shown in the study by~\cite{yoon2020chest} that except for more clarity in CT images, both modalities capture almost all the features associated with COVID-19. This fact has encouraged us to propose the same CNN architecture for the analysis of these two types. It is a well-known fact that deep learning models are highly data sensitive and limited data can suffer from over-fitting. It has been shown that as the number of parameters increase the model tries to remember the training data and give perfect results on hold-out test sets but fail to generalize on external data/real world scenarios~\cite{nguyen2021deep}. To address this issue of limited image availability, we have focussed on building a light-weight model with fewer trainable parameters. We also performed an experiment to see if the CNN model trained on CT images would perform better on CXR data, compared to when the model was trained on CXR data to understand the performance of the model trained on different image modality. This was done since more CT scan images are available for training compared to CXR images and a model that can identify the GGO patches in a CT may be able to identify similar patches in a CXR as the CT scan images can identify the abnormalities captured by CXRs in a more well defined manner and with less ambiguity~\cite{yoon2020chest}.

To evaluate the performance of proposed CNN model, few state-of-the-art deep learning models, viz., VGG-16~\cite{simonyan2014very}, ResNet-50~\cite{he2016deep} and Inception-v3~\cite{szegedy2016rethinking}, pre-trained on ImageNet database, are used. VGG-16 model has 16 layers with weights, the fully connected layers from the model were replaced with four customized fully connected layers of 512, 128, 64 and 1 (3 in case of three-class classification) nodes in each layer respectively. ResNet-50 has Conv blocks and Identity blocks arranged alternately to form 48 layers along with 1 ‘Maxpool’ and 1 ‘Avgpool’ layer. An output layer with 1 (3 in case of three-class classification) node was added to the model before training. Inception-v3 has multiple symmetric and asymmetric building blocks, including convolutions, average pooling, max pooling, and fully connected layers. Like in VGG-16, four additional fully connected layers were added to the model before training. Learning rate was set to 0.0001 initially for all three models and set to reduce by 0.3 if no improvement in validation loss was observed for 2 epochs. Activation functions were similar to that of the proposed CNN. Performance comparison is carried out for various parameters, viz., accuracy, precision, recall, F1-score, time taken to train the model, number of model parameters, etc. Precision, recall and F1-score are reported per class and accuracy is computed considering all classes. Accuracy is calculated as TP / TP +0.5*(FP + FN), where TP, FP and FN correspond to the number of true positives, false positives and false negatives, respectively. Recall (Sensitivity) gives the proportion of correct positive results and is defined as the ratio of true positives to all positive cases, TP/(TP+FN). Precision (Positive Predictive Value - PPV) gives the proportion of positively classified cases that are truly positive: TP/(TP+FP) and is a measure of how relevant a positive result is. F1-score is calculated the same way as accuracy but for each class separately.

\section{Implementation Details}
The model was trained on 4 GeForce GTX 1080 Ti GPUs and the time taken for training the CT model was\((\sim62)\) hours for 30 epochs and \(\sim10\) hours for 20 epochs in the case of CXR model. The optimizer used was Adam with an initial learning rate set to \(5e-5\) and decay rate of first and second moments were set to the default values of 0.9 and 0.999, respectively. The learning rate was set to reduce by 0.3 if no improvement in validation loss was observed for 2 epochs. With the initial learning rate set to the default value of Keras API (= 0.001), the model exhibited very low training and validation accuracy \((\sim0.46)\) and did not improve further. So, the learning rate was heuristically reduced till at \(5e-5\), the model performance significantly improved to 0.97. Further, reduction to \(5e-6\) did not show any improvement in the performance. Our analysis revealed that the lower and upper bounds of the optimal learning rate for this system lies between \(5e-3\) and \(5e-6\). The versions of software and framework used are given in supplementary file, S1.

\section{Results and Discussion}
\subsection{Chest CT Analysis}
\begin{figure}[!t]
\centering
\includegraphics[width=3.5in]{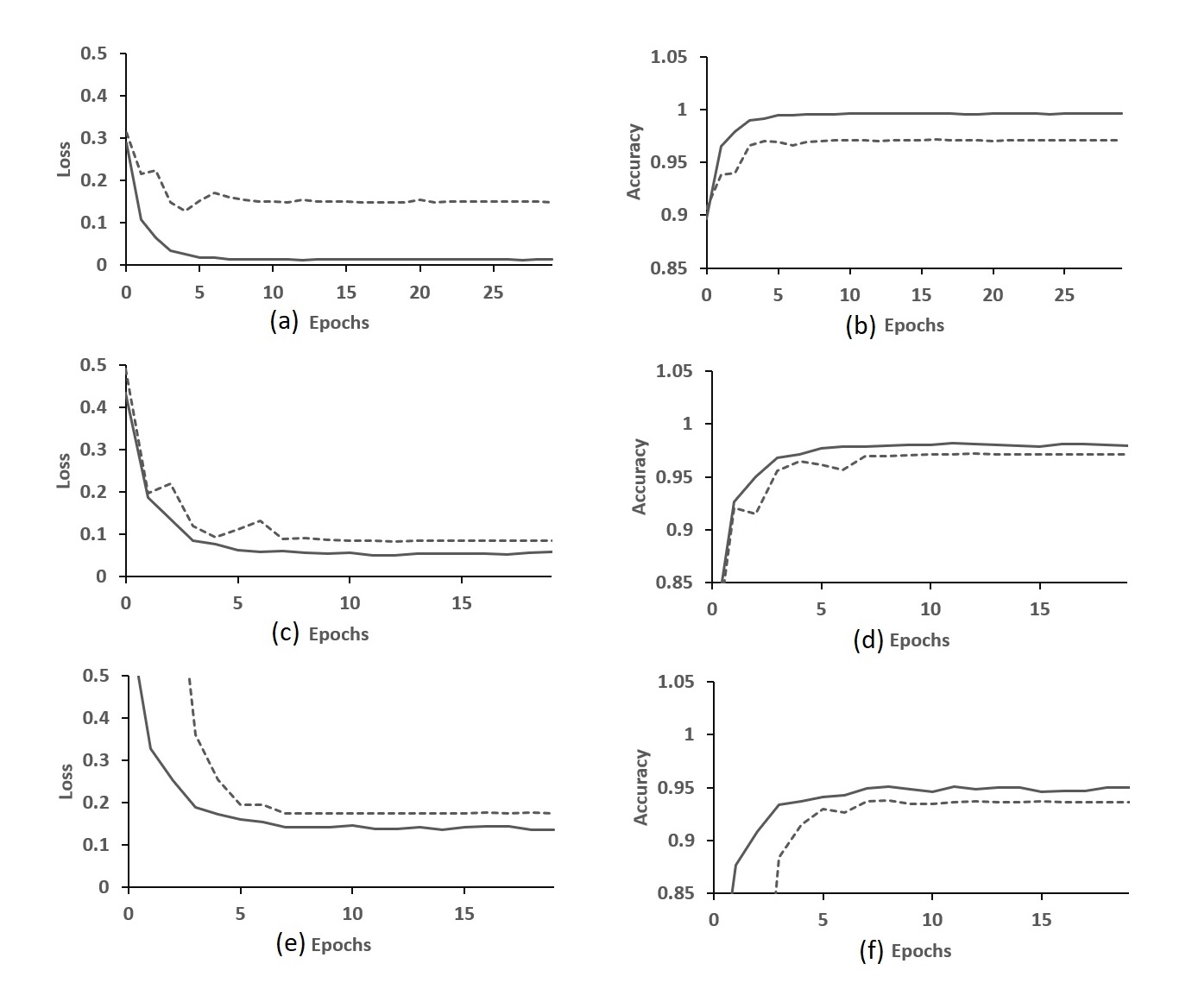}
\caption{Accuracy (solid line) and Loss Curves (dashed line) of training of the CNN model are shown. (a) and (b) on CT scan images, (c) and (d) of binary CXR classification, (e)and (f) of three-class CXR classification is shown.}
\label{fig_2}
\end{figure}

Performance evaluation of the proposed CNN model is carried out in two ways: (i) comparison with four state-of-the-art DL models, and (ii) on hold-out test set from COVIDx-CT dataset. Results of the analysis are summarised in Table~\ref{table_5}. The CNN model was trained for 30 epochs on COVIDx-CT dataset and training and validation accuracies converged to 0.99 and 0.96 respectively. For the CNN model, accuracy and loss curves (shown in Fig.~\ref{fig_2} a) and b)) plateau after 7 epochs showing that the model stabilizes without over-fitting. It is observed that the three DL models (except EfficientNetB7 
\((\sim23\%)\) achieved high accuracy by 3 epochs. ResNet-50 outperformed with precision and recall of 0.98 for the Covid class, followed by the CNN model with comparable values (0.96 and 0.94 respectively) for much fewer number of training parameters. VGG-16 (0.89) and Inception-v3 (0.82 and 0.84 respectively) had low precision and recall scores compared to CNN model.

\begin{table}
\renewcommand{\arraystretch}{1.3}
\caption{Performance evaluation of CNN and other DL models on COVIDx-CT test data (Inc-v3 : Inception-v3. Acc : Accuracy)}\label{table_5}
\begin{tabular}{|c|c|c|c|c|c|}
    \hline
    
    \bfseries Model & \bfseries  Class     & \bfseries  Precision    & \bfseries Recall      & \bfseries F1-score   & \bfseries Acc   \\
    \hline
    \multirow{3}{*}{\bfseries CNN} & Covid      & 0.96        & 0.94          & 0.95     &  \multirow{3}{*}{96\%}\\ 
    \cline{2-5}
    &    Normal     & 0.96        & 0.98          & 0.97             &    \\
    \cline{2-5}
    &    Pneumonia  & 0.99        & 0.97          & 0.98      &    \\
    \hline
    \multirow{3}{*}{\bfseries VGG-16} & Covid &  0.89          & 0.89          & 0.89     & \multirow{3}{*}{94\%}\\
    \cline{2-5}
    & Normal & 0.96          & 0.96          & 0.96  & \\
    \cline{2-5}
    & Pneumonia & 0.94          & 0.94          & 0.94   &   \\
    \hline
    \multirow{3}{*}{\bfseries ResNet-50} & Covid &  0.98          & 0.98          & 0.98     & \multirow{3}{*}{99\%}\\
    \cline{2-5}
    & Normal & 0.99          & 0.99          & 0.99  & \\
    \cline{2-5}
    & Pneumonia & 0.99          & 1.00          & 0.99   &   \\
    \hline
    \multirow{3}{*}{\bfseries Inc-v3} & Covid &  0.82          & 0.84          & 0.83     & \multirow{3}{*}{90\%}\\
    \cline{2-5}
    & Normal & 0.96          & 0.95          & 0.95  & \\
    \cline{2-5}
    & Pneumonia & 0.89          & 0.88          & 0.88   &   \\
    \hline
    \end{tabular} 
     \end{table}

\subsection{Chest X-ray Analysis}


\subsubsection{Binary Classification}
\begin{figure}[!t]
\centering
\includegraphics[width=3.0in]{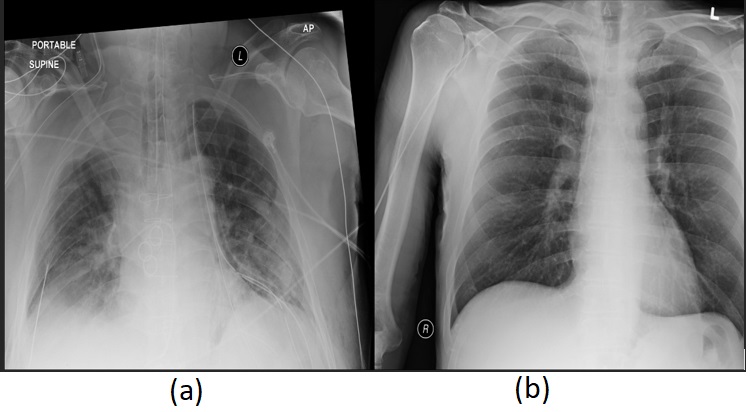}
\caption{The image (a) is that of non-COVID pneumonia, and (b) is of normal, that have been incorrectly predicted as belonging to COVID class.}
\label{fig_3}
\end{figure}

We now discuss the results on using the CNN model with the same architecture (Fig.~\ref{fig_1}) as used for CT scan image analysis, for binary and 3-class classification of CXR images, as for both the tasks almost same images were used. The binary classification was performed to analyse the model’s capability to differentiate between COVID-19 and non-Covid patterns in the images but the differences in the patterns between normal, non-Covid pneumonia and COVID-19 pneumonia can be verified only by a three-class classification. We first discuss the results of our proposed CNN model for binary classification of chest X-ray images given in Table~\ref{table_2}. The CNN model was trained on CXR images for 20 epochs, while the pre-trained VGG-16, ResNet-50 and Inception-v3 models were trained for 10 epochs, till convergence was achieved. This was confirmed from the training loss and accuracy curves in Fig.~\ref{fig_2} c) and d). From Table~\ref{table_6}, test accuracy value of the lightweight CNN model is comparable to the DL models, VGG-16 and ResNet-50, while the performance of Inception-v3 is very low. It is worth mentioning that time taken to train the CNN model is much lower \((\sim10\) hours) while for other models it is about \begin{math}\sim 4 \times\end{math} higher. The confusion matrix for binary classification of X-ray images is given in Table~\ref{table_7}. It is observed that a total of 14 incorrect classifications using CNN model (12 COVID-19 cases misclassified as non-Covid, and 2 non-Covid cases misclassified as COVID-19). To understand what went wrong in the prediction, the two non-Covid images predicted as covid class by the CNN model given in Fig.~\ref{fig_3} are manually analysed. Based on ground truth, first image (Fig.~\ref{fig_3}a) belongs to pneumonia class and second image (Fig.~\ref{fig_3}b) to normal class. The image in Fig.~\ref{fig_3}a is also misclassified by VGG-16 model. Both the images appear to have opacities in the lung areas, which may probably be because of noise in the image that led to misclassification. Asymptomatic patients may not have abnormalities in the lung images and hence may look like a normal image and some abnormalities may have very similar manifestations as that of pneumonia images which may not be differentiated by our binary classifier. This might cause such images getting predicted as non-Covid class by the model. Our analysis to understand the poor performance of Inception-v3 revealed that many non-Covid cases are predicted as covid. This indicates the inability of the model to differentiate between pneumonia and covid features (since the non-Covid class include both normal and non-Covid pneumonia images). Precision, recall and F1-score values of lightweight CNN model in Table~\ref{table_8} is comparable to the DL models VGG-16 and ResNet-50 in distinguishing between COVID-19 and non-Covid cases. However, the performance of Inception-v3 is very poor with low recall and F1-score values for the non-Covid class. We also experimented by using the model trained on CT images to classify CXR, but the results were not promising as reported in~\cite{panwar2020deep}.

\begin{table}
\renewcommand{\arraystretch}{1.3}
\caption{Performance comparison of the proposed CNN with VGG-16, ResNet-50 and InceptionNetV3 models for binary classification of chest X-ray images. Time taken for training the model (in hours) is given in brackets}
\label{table_6}
\centering
\begin{tabular}{|c|c|c|c|}
\hline
\bfseries Model/Acc &  \bfseries Train & \bfseries Validation & \bfseries Testing \\
\hline
    CNN   & 97.97 (9)  & 97.16      & 96.50   \\
\hline
    VGG-16     & 99.54 (34)  & 96.62     & 96.50     \\
\hline
    ResNet-50    & 1.00 (45) & 98.13     & 96.75      \\
\hline
 Inception-v3    & 96.07 (12)  & 55.63     & 55.50       \\
\hline
\end{tabular}
\end{table}

\begin{table}
\renewcommand{\arraystretch}{1.3}
\caption{Confusion Matrix for binary classification of CXR using the four models}
\label{table_7}
\centering
\begin{tabular}{|c|c|c|c|c|}
\hline
& \multicolumn{2}{|c|}{\bfseries CNN}  &  \multicolumn{2}{|c|}{\bfseries VGG-16} \\
\hline
 & non-Covid & COVID-19 & non-Covid & COVID-19  \\
\hline
 non-Covid &   198   & 2   & 198      & 2    \\
\hline
   COVID-19 & 12     & 188   & 12     & 188     \\
\hline
   \end{tabular}
   \begin{tabular}{|c|c|c|c|c|}
\hline
& \multicolumn{2}{|c|}{\bfseries ResNet-50}  &  \multicolumn{2}{|c|}{\bfseries Inception-v3} \\
\hline
 & non-Covid & COVID-19 & non-Covid & COVID-19  \\
\hline
 non-Covid &   199   & 1  & 44      & 156    \\
\hline
   COVID-19 & 12     & 188   & 22     & 178     \\
\hline
   \end{tabular}

\end{table}

\begin{table}
\renewcommand{\arraystretch}{1.3}
\caption{Performance comparison of the proposed CNN with VGG-16, ResNet-50 and Inception-v3 models for binary classification of CXR images}\label{table_8}
\begin{tabular}{|c|c|c|c|c|}
    \hline
    
    \bfseries Model & \bfseries  Class     & \bfseries  Precision    & \bfseries Recall      & \bfseries F1-score    \\
    \hline
    \multirow{2}{*}{\bfseries CNN} & non-Covid      & 0.94        & 0.99          & 0.97     \\ 
    \cline{2-5}
    &    COVID-19     & 0.99        & 0.94          & 0.96                 \\
    
    \hline
    \multirow{2}{*}{\bfseries VGG-16} & non-Covid &  0.94          & 0.99          & 0.97     \\
    \cline{2-5}
    & COVID-19 & 0.99          & 0.94          & 0.96   \\
    \hline
    \multirow{2}{*}{\bfseries ResNet-50} & non-Covid &  0.94          & 0.99          & 0.97     \\
    \cline{2-5}
    & COVID-19 & 0.99          & 0.94          & 0.97   \\
    \hline
    \multirow{2}{*}{\bfseries Inception-v3} & non-Covid &  0.67          & 0.22          & 0.33     \\
    \cline{2-5}
    & COVID-19 & 0.53          & 0.89          & 0.67   \\
    \hline
    \end{tabular} 
     \end{table}

\subsubsection{Multi-class Classification}
The training, validation and test accuracies of all the models for three-class classification using CXR images is given in Table~\ref{table_9} and the training and validation accuracy and loss curves are given in Fig.~\ref{fig_2} e) and f). From the confusion matrix in Table~\ref{table_10} we observe that the number of mis-predictions for the COVID-19 class by the CNN model is the least (5) compared to VGG-16 (8) and ResNet-50 (9) and Inception-v3 models (14). As in the case of binary classification, performance of CNN model is comparable to VGG-16 and ResNet-50 models. To understand why the CNN model failed in correctly predicting the five images in COVID-19 class visualization of the predictions are generated using heatmaps, the discussion of which is given in the next section. In multiclass classification, training time taken for the CNN model for 20 epochs is \((\sim10\) hours). The precision, recall and F1-score in Table~\ref{table_11} also indicates that the proposed CNN model is sufficient for this task. The precision, recall and F1-score for COVID-19 class were all highest for the simple CNN model, highest for ResNet-50, in the case of Pneumonia. VGG-16 had highest precision in the case of Normal class, recall was highest in ResNet-50 model. From the above results it can be deduced that, for building a clinical tool for detecting COVID-19, a simple CNN model is a better choice compared to complex deep learning models. The number of parameters in the CNN model is \(\sim2\) million depending upon the input dimensions used whereas VGG-16 has 138 million parameters, ResNet-50 has over 23 million parameters and Inception-v3 has 24 million parameters. This makes our model suitable for deploying in clinical settings as it is very ‘light weight’ and the training time for this simple model is much smaller compared to other state of the art deep learning models.

\begin{table}
\renewcommand{\arraystretch}{1.3}
\caption{Training, validation and test accuracies of the models for three-class classifications of CXR images is given. Time taken for training the model (in hours) is given in brackets}
\label{table_9}
\centering
\begin{tabular}{|c|c|c|c|}
\hline
\bfseries Model/Acc &  \bfseries Train & \bfseries Validation & \bfseries Testing \\
\hline
    CNN   & 95.02 (10)  & 93.61      & 94.24   \\
\hline
    VGG-16     & 93.50 (33)  & 92.45     & 94.24     \\
\hline
    ResNet-50    & 99.97 (45) & 96.30     & 96.75      \\
\hline
 Inception-v3    & 92.20 (14)  & 56.59     & 59.25       \\
\hline
\end{tabular}
\end{table}

\begin{table}
\renewcommand{\arraystretch}{1.3}
\caption{Confusion Matrix for the three-class classification of CXR using the four models}
\label{table_10}
\centering
\begin{tabular}{|c|c|c|c|}
\hline
& \multicolumn{3}{|c|}{\bfseries CNN}   \\
\hline
 & Covid-19 & Normal & Pneumonia \\
\hline
 COVID-19 &   195   & 2   & 3        \\
\hline
   Normal & 0     & 94   & 6         \\
\hline
Pneumonia & 2     & 10   & 88         \\
\hline
   \end{tabular}
   \begin{tabular}{|c|c|c|c|}
\hline
&   \multicolumn{3}{|c|}{\bfseries VGG-16} \\
\hline
 & Covid-19 & Normal & Pneumonia \\
\hline
 COVID-19 &   192   & 5   & 3       \\
\hline
   Normal & 2     & 95   & 3         \\
\hline
Pneumonia & 3     & 7   & 90         \\
\hline
   \end{tabular}

   \begin{tabular}{|c|c|c|c|}
\hline
& \multicolumn{3}{|c|}{\bfseries ResNet-50}   \\
\hline
 & Covid-19 & Normal & Pneumonia \\
\hline
 COVID-19 &   196   & 3  & 1      \\
\hline
   Normal & 0     & 98   & 2       \\
\hline
Pneumonia & 0     & 7   & 93       \\
\hline

   \end{tabular}
\begin{tabular}{|c|c|c|c|}
\hline
&   \multicolumn{3}{|c|}{\bfseries Inception-v3} \\
\hline
 & Covid-19 & Normal & Pneumonia \\
\hline
 COVID-19 &   186   & 13  & 1       \\
\hline
   Normal & 71     & 28   & 1      \\
\hline
Pneumonia & 75     & 2   & 23       \\
\hline

   \end{tabular}

\end{table}

\begin{table}
\renewcommand{\arraystretch}{1.3}
\caption{Performance evaluation of CNN and other DL models on COVIDx-CT test data (Inc-v3 : Inception-v3}\label{table_11}
\begin{tabular}{|c|c|c|c|c|}
    \hline
    
    \bfseries Model & \bfseries  Class     & \bfseries  Precision    & \bfseries Recall      & \bfseries F1-score     \\
    \hline
    \multirow{3}{*}{\bfseries CNN} & Covid      & 0.99        & 0.97          & 0.98     \\ 
    \cline{2-5}
    &    Normal     & 0.89        & 0.94          & 0.91                \\
    \cline{2-5}
    &    Pneumonia  & 0.91        & 0.88          & 0.89          \\
    \hline
    \multirow{3}{*}{\bfseries VGG-16} & Covid &  0.97          & 0.96          & 0.97     \\
    \cline{2-5}
    & Normal & 0.89          & 0.95          & 0.92   \\
    \cline{2-5}
    & Pneumonia & 0.94          & 0.90          & 0.92     \\
    \hline
    \multirow{3}{*}{\bfseries ResNet-50} & Covid &  1.00          & 0.98          & 0.99     \\
    \cline{2-5}
    & Normal & 0.91          & 0.98          & 0.94   \\
    \cline{2-5}
    & Pneumonia & 0.97          & 0.93          & 0.95     \\
    \hline
    \multirow{3}{*}{\bfseries Inc-v3} & Covid &  0.56          & 0.93          & 0.70     \\
    \cline{2-5}
    & Normal & 0.65          & 0.28          & 0.39   \\
    \cline{2-5}
    & Pneumonia & 0.92          & 0.23          & 0.37     \\
    \hline
    \end{tabular} 
     \end{table}

\section{Visualization of the results}
To identify regions in the lungs that are characteristic of COVID-19 infection, Gradient-weighted Class Activation Mapping (Grad-CAM) technique is used. Grad-CAM produces heatmaps that highlight regions involved in arriving at the predictions made. This aids in visual inspection of the prediction, just like an expert radiologist would while making the decision. Grad-CAM uses the gradients of the output of a model that flow into the final convolution layer to produce a localization map highlighting the important regions in the image that contributed towards the prediction~\cite{selvaraju2017grad}. These gradients are then subjected to global average pooling to obtain weights for the target class. The weighted combination of the activation map is extracted and subjected to ‘Relu’ operation to generate a heatmap of the same size as that of the feature map obtained from the convolutional layer. This heatmap when overlayed on the original image highlights the pixels that contributed to the prediction. The heatmap can help the diagnostician in understanding the reliability of model’s predictions and can be a useful tool in triaging the patients based on the severity of the infection. 

\begin{figure}[!t]
\centering
\includegraphics[width=2.0in]{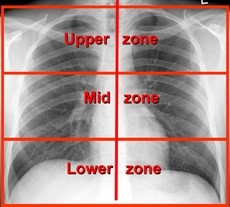}
\caption{The 6 lung zones obtained by dividing a frontal CXR is shown. Image taken from [6].}
\label{fig_4}
\end{figure}

As shown in Fig.~\ref{fig_4}, the frontal chest radiograph can be divided into 3 zones per lung, i.e., a total of 6 zones. According to the proposed reporting protocol for grading lung disease related to COVID-19 on frontal CXR by~\cite{litmanovich2020review}, if there are opacities in 1-2 lung zones the disease is mild, if found in 3-4 zones, moderate, and if more than 4 zones have opacities it is a severe case of COVID-19.

\begin{figure}
\centering
\includegraphics[width=3.0in]{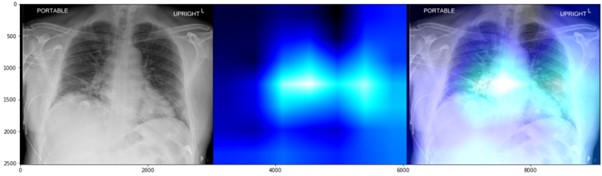}
\caption{Grad-CAM visualization of the image of a patient who was diagnosed to have COVID-19 and model gave the correct prediction}
\label{fig_5}
\end{figure}

\begin{figure}
\centering
\includegraphics[width=3.0in]{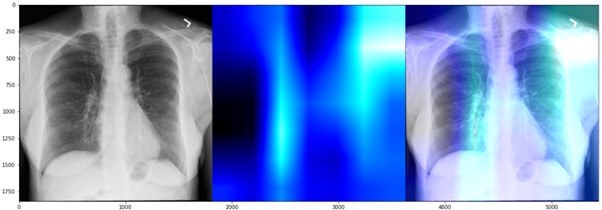}
\caption{Grad-CAM visualization of the image of a patient who was diagnosed to have COVID-19 but model predicted it to be non-Covid}
\label{fig_6}
\end{figure}

\begin{figure}
\centering
\includegraphics[width=3.0in]{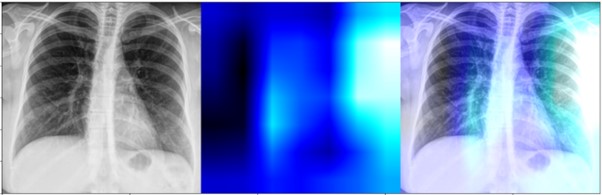}
\caption{Grad-CAM visualization of the image of a patient who was diagnosed to have COVID-19 but model predicted it to be non-Covid}
\label{fig_7}
\end{figure}

\begin{figure}
\centering
\includegraphics[width=3.0in]{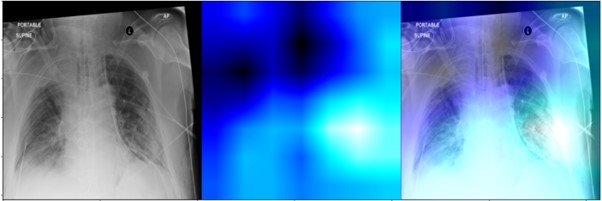}
\caption{Grad-CAM visualization of the image of a patient who was negative to COVID-19 but model predicted it to be COVID-19}
\label{fig_8}
\end{figure}

\begin{figure}
\centering
\includegraphics[width=3.0in]{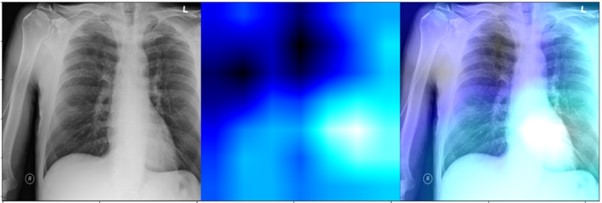}
\caption{Grad-CAM visualization of the image of a patient who was negative to COVID-19 but model predicted it to be COVID-19}
\label{fig_9}
\end{figure}

\begin{figure}[!t]
\centering
\includegraphics[width=3.0in]{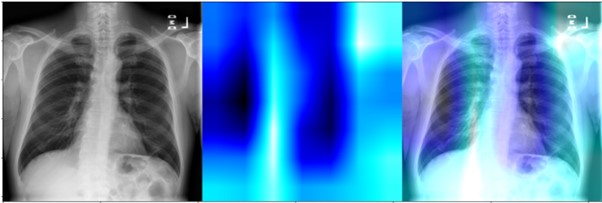}
\caption{Grad-CAM visualization of the image of a patient who was negative to COVID-19 and model model gave the correct prediction}
\label{fig_10}
\end{figure}

In Figures~\ref{fig_5} -~\ref{fig_10} are shown a few representative images from the test set that correspond to correct and incorrect predictions. The objective of the exercise is to understand how the model arrived at correct predictions, that is, what areas in the image are identified as covid associated patterns in the lungs, and why the model failed in some cases. In Fig.~\ref{fig_5}, CXR image of a COVID-19 patient (left panel), heatmap extracted from the last convolutional layer of the CNN model (middle panel) and heatmap overlaid on the original image (right panel) is shown. The CXR in the figure has opacities in at least 3 zones in the lungs and is clearly a positive case of COVID-19. It is evident from the heatmap that the blobs are covering the correct regions of the lungs with opacities and in the overlaid image these regions are clearly highlighted. In Figures~\ref{fig_6} and~\ref{fig_7} CXR images of COVID-19 cases are shown for which the model made incorrect predictions. No opacities are seen in the CXR images and the heatmap extracted also do not have any blobs covering the lung region and hence the ‘negative’ prediction. The probable explanation of no COVID features in the CXR image may be that the image was taken during early days of infection or from an asymptomatic COVID-19 patient. This is a limitation of the approach since analysis of image data can identify the disease only when the lungs are infected. Thus, other clinical features of the disease are also important and must be taken into consideration along with the chest image data. Figures~\ref{fig_8} and~\ref{fig_9} are CXR images of non-Covid cases, but model predicted these as COVID-19. We clear see opacities in the CXR images which are highlighted as blobs in the heatmaps. Since the images in the negative set also include other pneumonia cases, apart from normal cases, it is probable that severity of lung infections in some cases of pneumonia may exhibit accumulation of fluid in the lungs which may be mistaken with COVID-like features. Fig.~\ref{fig_10} is a CXR image of a non-Covid case, and the model prediction agrees with the ground truth. The image is clear of any opacities and no blobs are seen in the extracted heatmap. From the analysis we observe that the model predictions are reliable when the patients’ lungs are infected and thus can be a useful tool in monitoring the progress of the disease. However, since during early phase of the disease or if the patient is asymptomatic or has infection only in the nasal and throat regions, chest radiographs may not be very useful in the detection of the disease.   

\section{Conclusion}
In this work we present the use of CXR and CT imaging to differentiate the patterns of COVID-19 infections from other lung infections. A lightweight CNN architecture is proposed because of their pattern recognition ability for the analysis of both chest CXR and CT scan images. Here we show that the performance of CXR image analysis is comparable to CT scan analysis making it possible to extend to larger populations and at patient bedside. To analyse the prediction result of the CNN model, a visualization technique called Grad-CAM is implemented that highlights pixels corresponding to Ground Glass Opacities and consolidations in the lung areas, which are characteristics of COVID-19. The proposed CNN model has \(\sim2M\)  parameters, much fewer most deep learning models, but with comparable performance. Thus, it can be easily deployed to smart devices for clinical applications. Since negative CT result does not rule out COVID-19 infection, there has been some concern in using CT scans for the detection of COVID-19 by most radiological societies to avoid unnecessary exposure to radiation. X-rays, on the other hand has much lower exposure to radiation, making it safe to use for fast triaging of patients in a pandemic situation. Another concern has been dependence of the deep learning models on the training dataset size, since learning using very deep models on limited datasets can lead to overfitting and make the model less generalizable. Thus, there is a clear need for developing large collection of CXR and CT image datasets from different population groups which would be of great help in model development along with continuous learning.


%



  \section*{Acknowledgment}

The authors would like to thank IHub-Data, IIIT-Hyderabad for the financial support provided.




\bibliographystyle{IEEEtran}
\bibliography{ieee}

%
%
%

\vfill




\end{document}